\documentclass[10pt,letterpaper]{article}
\usepackage[top=0.85in,left=1in,footskip=0.75in,marginparwidth=2in]{geometry}

\usepackage[utf8]{inputenc}

\usepackage{cite} 

\usepackage{nameref,hyperref}

\usepackage{microtype}
\DisableLigatures[f]{encoding = *, family = * }

\raggedright
\usepackage{float}  

\usepackage{amsmath}      
\usepackage{graphicx}     
\usepackage{amssymb}      
\usepackage{hyperref}     
\usepackage{makecell} 

\usepackage{changepage}

\usepackage[aboveskip=1pt,labelfont=bf,labelsep=period,singlelinecheck=off]{caption}

\makeatletter
\renewcommand{\@biblabel}[1]{\quad#1.}
\makeatother

\usepackage{lastpage,fancyhdr,graphicx}
\usepackage{epstopdf}
\fancyheadoffset[L]{2.25in}
\fancyfootoffset[L]{2.25in}

\usepackage{color}

\definecolor{Gray}{gray}{.25}

\usepackage{graphicx}

\usepackage{sidecap}

\usepackage{wrapfig}
\usepackage[pscoord]{eso-pic}
\usepackage[fulladjust]{marginnote}
\reversemarginpar

\begin{document}
\vspace*{0.35in}

\begin{flushleft}
{\Large
\textbf\newline{\ Specific Nucleic Acid Detection Using a Nanoparticle Hybridization Assay}
}
\newline
\\
Arwa A. Aldakheel\textsuperscript{1},
Christopher B. Raub\textsuperscript{1},
Hieu Bui\textsuperscript{2},

\bigskip
\bf{1} Departments of Biomedical Engineering, The Catholic University of America, 620 Michigan Avenue N.E.,
Washington, D.C. 20064, USA
\\
\bf{2} Departments of Electrical Engineering and Computer Science, The Catholic University of America, 620
Michigan Avenue N.E., Washington, D.C. 20064, USA
\\
\bigskip

* raubc@cua.edu

\end{flushleft}

\section*{Abstract}
Simple methods to detect biomolecules including specific nucleic acid sequences have received renewed attention since the Severe Acute Respiratory Syndrome Coronavirus 2 (SARS-CoV-2) virus pandemic. Notably, biomolecule detection that uses some form of signal amplification will have some form of amplification-related error, which in the polymerase chain reaction involves mispriming and subsequent signal amplification in the no template control, ultimately providing a limit of detection. To demonstrate the feasibility of the detection of a DNA target sequence without molecular or chemical signal amplification that avoids amplification errors, a gold nanoparticle aggregation assay was developed and tested. Two primers bracketing a 94 base pair target sequence from SARS-CoV-2 were conjugated to 10 nm diameter gold nanoparticles by the salt aging method, with conjugation and primer-target hybridization confirmed by agarose gel electrophoresis and nanospectrophotometry. Upon mixing of both conjugated nanoparticles with target, a surface plasmon resonance shift of 6 nm was observed, and lower electrophoretic mobility of a band containing both DNA fluorescence and gold absorption signals. This did not occur in the presence of a control DNA molecule of the same size and composition as the target but with a randomly scrambled base position. Nanoparticle tracking at 30 frames per second using a sensitive darkfield microscope revealed a lower measured diffusion coefficient of scattering objects in the target mixture than in the control mixture or with bare gold nanoparticles. Stochastic diffusion and reaction models confirmed the accuracy of the diffusion coefficient estimation from nanoparticle tracks, and the presence of dimers, trimers, and tetramers, depending on the stoichiometric amount of target DNA mixed with conjugated detection nanoparticles. These findings support the future development of nanoparticle assays based on freely diffusing nanoparticles and suggest that real-time tracking of nanoparticles could be used to detect even single-target hybridization events.


\section*{Introduction}
Molecular diagnostic assays detect specific nucleic acid sequences for basic biomedical research and clinical testing. Pathogens, cancer cells, and normal somatic cells with genetic mutations associated with hereditary risk for disease all leave fragmentary genetic material in the blood and other biofluids or in situ, accessible through clinical collection techniques. After steps of nucleic acid isolation and purification, polymerase chain reaction is a powerful and common approach to detect specific sequences. Next-generation sequencing, digital PCR, isothermal nucleic acid amplification, and fluorescence in situ hybridization are other approaches to detect specific targets \cite{stenson2014human, katsanis2013molecular}. Most of these approaches require molecular amplification of the target sequence, and all require hybridization of regions of the target sequence to designed complementary oligomers or primers.
The recent pandemic caused by a coronavirus, SARS-CoV-2, created an immediate need for rapid, widespread, and accurate testing of sick patients, in hospitals, medical centers, and at home \cite{younes2020challenges, cheng2020diagnostic}. As part of the pandemic response, clinical diagnostic centers upscaled the use of reverse transcription-polymerase chain reaction to detect SARS-CoV-2 specific genetic sequences \cite{liu2020positive, ahmed2022minimizing,mak2020evaluation}. Another critical part of pandemic management was use of over-the-counter testing kits based on SARS-CoV-2 protein antigen recognition from mucus collected by nasal swabs, allowing individuals to test themselves and family members at home \cite{li2021immunologic,di2021detection,coste2021comparison}. While the presence of appreciable amounts of protein antigen indicates transcription of viral genetic material in the present or recent past, the presence of virus-specific nucleic acids is potentially a more sensitive indicator of viral exposure, with the amount nucleic acid targets detected dependent on multiple factors including time from first exposure, repeated exposures, presence and strength of immune response, and the timecourse of viral load in an active or recent infection \cite{ahmed2022minimizing,brummer2021accuracy,tombuloglu2021development}.
Sensitive molecular tests for specific DNA sequences are useful beyond testing for viral exposure, however. The genetic mutations corresponding to neoplasms and precancerous genetic changes in tissues may not be present in high abundance in easily accessible locations, for example in blood, saliva or mucus. Also genetic heterogeneity in solid tumors indicates that potentially important genetic mutations that drive acquired resistance to targeted therapies may also not be present at high levels in a given biopsy or as cell-free DNA or within exosomes in circulating blood \cite{gautam2023mucins, fan2010detection, cho2010micrornas, heneghan2010systemic}. Therefore, further development of sensitive molecular assays for specific target nucleic acid sequences is warranted for a number of clinical applications.
Nanoparticle aggregation based on binding to specific molecular targets is one recent innovation with the potential for new clinical diagnostic applications. Gold, silver, iron oxide, silica, and plastic nanoparticles may be decorated with antibodies, aptamers, or oligomers through specific conjugation chemistry to bind biomolecules leading to aggregation. Very often ensembles of aggregates are required to produce a visible color change in a mixture, or an electrophoretic mobility shift, meaning these assays do not readily detect small numbers of target molecules.
This work describes the detection of target DNA sequences using a novel approach of darkfield videomicroscopy followed by diffusion coefficient estimation through nanoparticle tracking. It was reasoned that darkfield detection of nanoparticle aggregation would be more sensitive than other methods, including spectrophotometry, dynamic light scattering, and colorimetric detection techniques, using simple, inexpensive equipment, small reagent volumes, and avoiding the need for target amplification. This study aims to demonstrate the feasibility of detecting small, freely diffusing nanoparticle aggregates induced by target presence, determine mechanisms leading to false positive detection, and separate ways to mitigate false positives. A nucleic acid target from SARS-CoV-2 was selected, primers were designed and optimized and conjugated to 10 nm diameter gold nanospheres. Then, nanoparticle aggregation in the presence of target and control sequences was assessed by nanospectrophotometry, gel electrophoresis, and nanoparticle tracking. Results indicate the feasibility of detecting specific nucleic acid sequences using the latter approach and suggest ways to reduce false positives associated with mis-hybridization and conjugation to non-target sequences.
\section*{Materials and Methods}
All single-stranded DNA oligonucleotides were procured from Integrated DNA Technologies (IDT, Coralville, IA) and designed to complement specific target sequences of SARS-CoV-2. The selected primer, target, and scrambled control sequences are detailed in (Table S1). Thiol-modified oligonucleotides were reduced using acetate buffer and Tris(2-carboxyethyl) phosphine hydrochloride (TCEP) from MilliporeSigma (Burlington, MA, USA) to facilitate conjugation with nanoparticles. Citrate buffer stabilized gold nanoparticles (AuNPs) with a diameter of 10 nm were purchased from MilliporeSigma.
\subsection*{1. Primer Design and Optimization}
Target sequences from the SARS-CoV-2 genome ( NC-045512.2 ) were analyzed for primer optimization around the 810 base pair RNA-dependent RNA polymerase (RdRp) region within open reading frame 1 (GenBank accession number MT072668.1), which is known to be highly conserved across different coronaviruses. Using PrimerBlast (NCBI), we initially generated 10 primers each spanning 30 nucleotides. Then each primer was analyzed with the primer optimization tool from IDT (details in supplement). Oligo A (forward) corresponded to gene position 9375–9404, while Oligo B (reverse of reverse complement) spanned gene position 9447–9418. Both primers targeted the same single strand of 73 bp-long target DNA. A DNA sequence with the same base composition as the target, but with bases in randomized order, was used as a control in experiments (‘scrambled control’). To facilitate conjugation and hybridization, a 10-mer poly T spacer was added between the thiol group and the hybridization region \cite{demers2000fluorescence, ochtrop2020recent}.
\subsection*{2. NUPACK oligomer analysis}
Primer target and self-hybridization efficiency were evaluated using NUPACK, which provided insights into the primer-primer and target-target interactions and their hybridization rates. Oligo A, oligo B, target, and scrambled sequences were all represented at 1 $\mu$M, and interactions were simulated. The melting temperatures (Tm's) and free energy of the structure of products were calculated and compared. All 10 primer options were analyzed in this way to select the pair with minimal primer-dimer formation, self-hybridization, and lowest free energy of the primer-target product (with high free energy of the primer-scrambled product). Among the 10 primers analyzed, two were chosen to serve as oligomers having met all acceptance criteria listed above.
\subsection*{3. Nanoparticle oligomer conjugation}
\begin{figure}[H] 


\includegraphics[width=\textwidth]{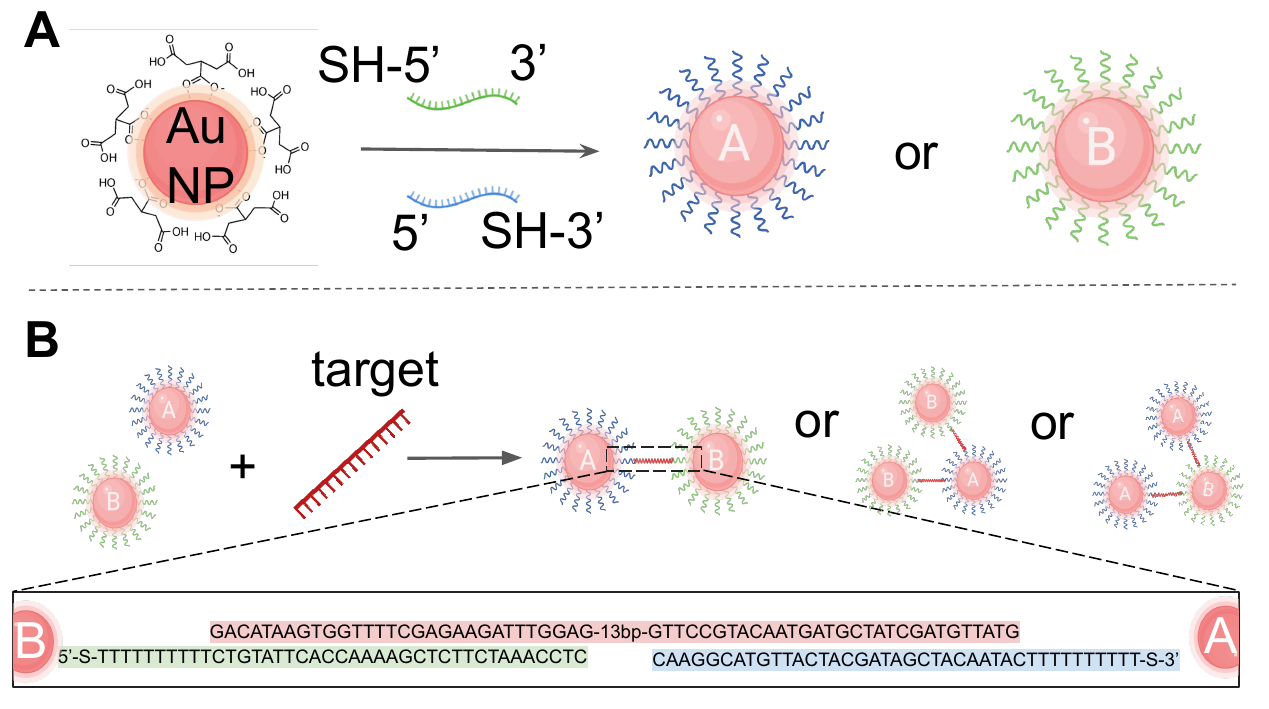}

\caption{\color{Gray} \textbf{Gold nanoparticle conjugates for DNA detection.} \textbf(A) Citrate-capped gold nanoparticles (AuNPs) were functionalized with thiol-modified oligonucleotides, sequence A (blue) and B (green). (B) Oligo-conjugated AuNPs bound a target DNA sequence to form A-B dimers, trimers, and potentially higher aggregates (not depicted). The target sequence (red) was complementary to SARS-CoV-2 genome.}

\label{fig1} 

\end{figure}

Conjugation of DNA to AuNPs (Figure \ref{fig1}A) was conducted following a modified version of a previously established protocol \cite{liu2006preparation}. Initially, thiolated oligonucleotides were reduced at room temperature for 1 hour utilizing a solution comprising 500 mM acetate buffer and 10 mM freshly prepared TCEP. The resulting reduced oligonucleotides, with a final concentration of 3 $\mu$M, were then combined with AuNPs. The mixture was allowed to incubate at room temperature for 16 hours to enable efficient conjugation between the DNA and AuNPs. Subsequently, solutions of 1 M NaCl and 500 mM Tris-acetate were incrementally added dropwise to the reaction mixture until reaching final concentrations of 3 mM NaCl and 5 mM Tris-acetate, respectively. Gentle agitation of the mixture was performed after each addition of salt to prevent agglomeration. Finally, the mixtures were incubated overnight at room temperature.

\subsection*{4. Spectrophotometry and gel electrophoresis}
Nanoparticles conjugated to oligomers A and B were added to target/scrambled control at concentrations of 3 $\mu M$ each. Also, target was added separately at 6 and 9 $\mu M$, for 2:1 and 3:1 stoichiometric ratios of target to oligomers A and B. UV-vis spectroscopy was used to evaluate DNA interaction with gold nanoparticles. A Nanodrop 2000 spectrophotometer was used to measure absorption in the 200–800 nm range. Oligomeric DNA absorbed light in the wavelength range 190-350 nm. Gold nanoparticles surface plasmon resonance peaks were assessed in the wavelength range 500-550 nm.
Agarose gel electrophoresis was performed to assess conjugation, hybridization and size of DNA-AuNPs complexes. Agarose at 2 (Bio-Rad, Hercules, CA, USA) in 1x Tris/acetic acid/EDTA (TAE) buffer and loading the gel with DNA-AuNPs complexes \cite{yao2015clicking}. After conjugation, the complexes were mixed with gel-loading dye and loaded onto the wells along with a DNA ladder. The gel was then run at 100 V for 30 minutes in 1x TAE buffer. Following electrophoresis, the gel was visualized under a UV transilluminator and with epi white illumination (ChemiDoc MP Gel Imager, Bio-Rad). The migration pattern of DNA-AuNPs was compared to that of free DNA samples and bare gold nanoparticles. 
\subsection*{5. Darkfield Videomicroscopy}
Darkfield microscopy was performed using a commercial inverted microscope (ECLIPSE TE 300, Nikon Corp., Melville, NY), with a 40x magnification/1.3 numerical aperture oil-immersion objective and high numerical aperture (1.0) condenser. The darkfield aperture position was manually adjusted for optimal darkfield detection and resolution. Video-speed, high resolution micrographs were collected using a sensitive microscope camera with a 2208x2208 array of pixels of dimension 4.54 $\mu m$ in a charge-coupled device sensor (Photometrics X1 camera, Thermo Scientific). Video framerate was 30 ms. The imaging chamber was constructed by placing an imaging spacer (0.15 mm) with double-sided adhesive between two coverslips (0.17 mm). The spacer and coverslips formed a cylindrical chamber for imaging nanoparticles and aggregates. Bare AuNPs, and target or scrambled DNA mixed with nanoparticles conjugated to oligomers A and B were added to these imaging chambers. Videos were collected over 300 frames/9 seconds, with multiple nanoparticles tracked per field of view/video. In total, n=20-25 nanoparticles were tracked per group, for the 3 groups. Nanoparticles were randomly selected for tracking by digitally zooming to a smaller region of interest containing the nanoparticle, which allowed for an adequate 30 ms framerate. Nanoparticles were tracked after image collection using the TrackMate plugin of Fiji \cite{schindelin2012fiji,tinevez2017trackmate}. 
\subsection*{6. Diffusion coefficient estimation}
Two methods were used to estimate the population diffusion coefficients from nanoparticle tracking data \cite{catipovic2013improving}. These methods are based on modeling the nanoparticles’ motion as random walks, in which the two-dimensional mean squared displacement (MSD, or $\langle r^2 \rangle$) is given by Equation \ref{eq:1}:
\begin{equation}
\langle r^2 \rangle = \langle x^2 \rangle + \langle y^2 \rangle = 4Dt
\label{eq:1}
\end{equation}
in which $\langle x^2 \rangle$ and $\langle y^2 \rangle$ represent the one-dimensional MSDs, $D$ denotes the diffusion coefficient, and t signifies the time between data points. In the first way (Method 1) the diffusion coefficients were found by fitting the distribution of particle lateral displacements between successive frames to a Gaussian curve. The variance, $Var(\Delta x)$, equals the mean squared displacement for a one-dimensional random walk, $\langle \Delta x^2 \rangle$, if the net displacement $\Delta x$ is not different from zero:
\begin{equation}
Var(\Delta x) = \langle \Delta x^2 \rangle - \langle \Delta x \rangle^2
\label{eq:2}
\end{equation}
which can be determined by the central value of the displacement histogram Gaussian fit. Under this condition, the variance is equal to 2Dt, for the one-dimensional case. For Method 2, particle lateral displacement data was used to calculate the MSD using the following formula:
\begin{equation}
\langle r^2(t) \rangle = \frac{1}{N_i}\sum_{i=1}^{N_i} \left[(x_i(t) - x_i(0_i))^2 + (y_i(t) - y_i(0_i))^2 \right]
\label{eq:3}
\end{equation}
where $N$ is the number of image frames,$\langle x^2 \rangle$ and $\langle y^2 \rangle$ are the x and y positions of particle $i$ at time $t$, and $x_i(t)$ and $y_i(t)$ are the initial x and y positions of particle $i$. From these calculations, the MSD was plotted for a range of time intervals corresponding to different frames in the image stack. In general, the MSD was greater when the time interval was greater. This relationship was plotted, and a best-fit line was determined, whose slope was the estimated diffusion coefficient in $\mu\text{m}^2/\text{s}$ \cite{catipovic2013improving}.

Finally, the Stokes-Einstein equation was used to calculate the diffusion coefficient directly from nominal nanoparticle and environmental parameters. The Stokes-Einstein equation relates the diffusion coefficient $D$ to particle size $a$, fluid viscosity $\eta$, and temperature $T$ through the equation \cite{edward1970molecular}: 

\begin{equation}
D = \frac{k_B T}{6 \pi \eta a}
\label{eq:4}
\end{equation}
Where $K_B$ is the Boltzmann constant, $\eta$ is the viscosity of the fluid, and $a$ is the radius of the particle. 
This equation assumes spherical symmetry, a good assumption for the commercially available spherical gold nanoparticles which had a measured diameter average of 10 nm. Viscosity was assumed to be that of water at 27°C. Later on, aggregates were modeled as both spheres or cylinders.
\subsection*{7. Nanoparticle Diffusion and Reaction Kinetic Stochastic Models}
Stochastic models were used to simulate the diffusion characteristics of gold nanoparticles, aggregates, and separately, their stoichiometry of dimers, trimers, and tetramers in the end products of the conjugated nanoparticle-target hybridization reaction. First, the Stokes-Einstein equation solution to the diffusion coefficients of spherical and cylindrical nanoparticles were plotted versus i) sphere radius, and ii) the radius of a cylinder of aspect ratio 2. The diffusion coefficient equation for a sphere was given in Equation \ref{eq:4}. For a cylinder, the diffusion coefficient equation is:
\begin{equation}
D = \frac{k_B T \ln(L/a) + 0.193)}{4 \pi \eta L}
\label{eq:5}
\end{equation}

where $L$ is the length of the particle (the length of our DNA) along with its radius $a$ \cite{rudyak2020kinetic,stellwagen2003unified}.
Then, to predict the Brownian motion of spherical and cylindrical nanoparticles, a stochastic model was created based on solving the Langevin equation, which represents a nanoparticle diffusing in fluid media and subjected to force from Brownian motion (molecular collisions) and a drag force:
\begin{equation}
m \frac{d^2x}{dt^2} = -F_{\text{Drag}} \frac{dx}{dt} + F_B
\label{eq:6}
\end{equation}

Where $x$ is the position of the particle,  $F_{\text{Drag}}$ is the drag force, that depends on particle shape, and $F_B$ is force due to Brownian motion, which depends on a random variable in the model \cite{michaelides2015brownian}. Nanoparticle displacement was updated by solving the Langevin equation for a given set of nanoparticle parameters and a time step of 0.03 seconds using a Matlab solver (ode23). The kinetic model allowed for multiple target molecules to bridge nanoparticle-oligomer conjugates, forming dimers, trimers, and tetramers. The rates of formation of these reaction products followed first-order reaction kinetics equations. A system of 12 ordinary differential equations representing the rate of change of all reactants and products was solved using a Matlab solver (ode15s), for a sufficiently long timespan and assumed equal reaction rates to allow the system to attain an equilibrium. Initial conditions assumed equal concentrations of conjugate A, B, and T, and then varied the stoichiometric ratio of T between 0.1 and 10. The final equilibrium amounts of dimers, trimers, and tetramers as a percent of total T were plotted for each stoichiometric ratio of T. 
\subsection*{8. Statistical analyses}
To evaluate the effects of target and scrambled DNA sequence presence on nanoparticle diffusion coefficient estimates, a one-factor ANOVA was performed on D values obtained using methods 1 and 2, with post hoc Tukey tests. The significance level was set at $p < 0.05$.

\section*{Results}
\subsection*{1. Primer and target sequences}
The final primer sequences and target DNA sequence from the SARS-CoV-2 genome are listed in Figure \ref{fig1} and Table S1. The scrambled control sequence is also listed in Table S1. The theoretical distance separating the centers of two hybridized nanoparticles was calculated to be 41.6 nm assuming a bottlebrush configuration of oligomers and hybridized target around the spheres. The hybridization region of each oligomer had a separation of 3.4 nm from the nanosphere surface based on a 10-T spacer.
\subsection*{2. Thermodynamic and structural DNA analyses (NUPACK)}
\begin{figure}[H] 
\caption{\color{Gray} \textbf{Gel electrophoresis and structural analysis of oligomers and target DNA.} \textbf{DNA} sequences in the absence of gold nanoparticles were mixed and run on an agarose gel, then stained with Ethidium Bromide to confirm reactant and product sizes. The bright band highlighted by an arrowhead occurs in the lane with mixed oligomer A, oligomer B, and target, but not with oligomers and scrambled DNA. (A,B) Oligomer secondary structures, and (C) the hybridized product have negative free energies (indicated by the symbol G). (D) Pair probabilities equal to 1 of oligomers A and B hybridizing with the target sequence, over the entire base index of the molecules.}
\renewcommand{\thefigure}{2}
\includegraphics[width=\textwidth]{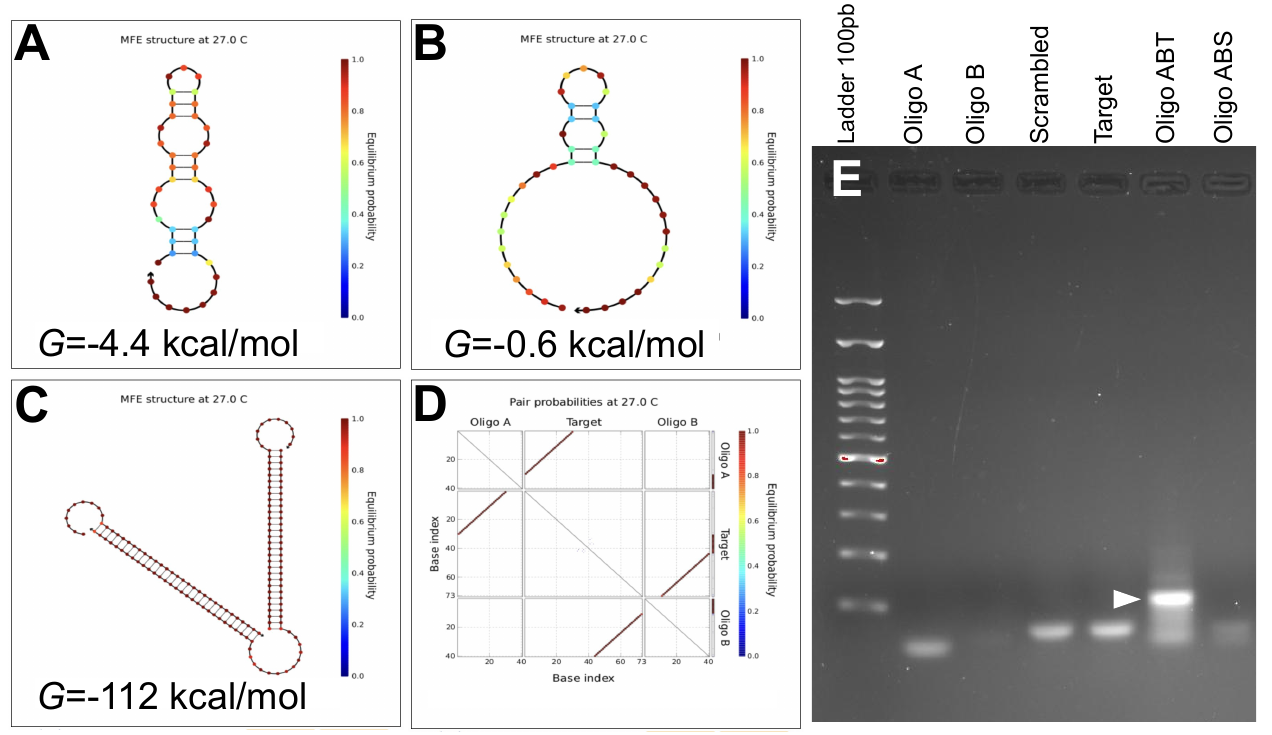}
\label{fig2} 
\end{figure}

Oligomer A (Figure \ref{fig2}A), B (Figure\ref{fig2}B), and the hybridized A-B-target (Figure\ref{fig2}C) secondary structures with minimum free energy contained stems, loops and hairpin structures. The free energy of the hybridization product was much lower than that of the self-interaction of either oligomer (-112 kcal / mol for the product vs. -4.4 and -0.6 kcal / mol for oligomers A and B, respectively). The hybridization product had a perfect base pairing in the hybridization regions (40 bp for each oligomer), with a 13 bp target region separating the two oligomer hybridization regions, and 10 bp poly-T tails (Figure \ref{fig2}C,D). The scrambled target sequence hybridized poorly with oligomer B alone, with one predominant predicted reaction product (containing one oligomer B and one scrambled molecule, Figure\ref{figS1}. Oligomer A had no predicted interaction with the scrambled product. DNA reactants and reaction mixtures produced bands with characteristic mobility in gel electrophoresis Figure\ref{fig2}E). The A-B-T mixture had the least mobility. The A-B-S mixture produced two bands, one with the same mobility as scrambled and one with slightly less mobility.

\subsection*{3. Target detection by gold nanoparticle aggregation}
\begin{figure}[H] 
\includegraphics[width=\textwidth]{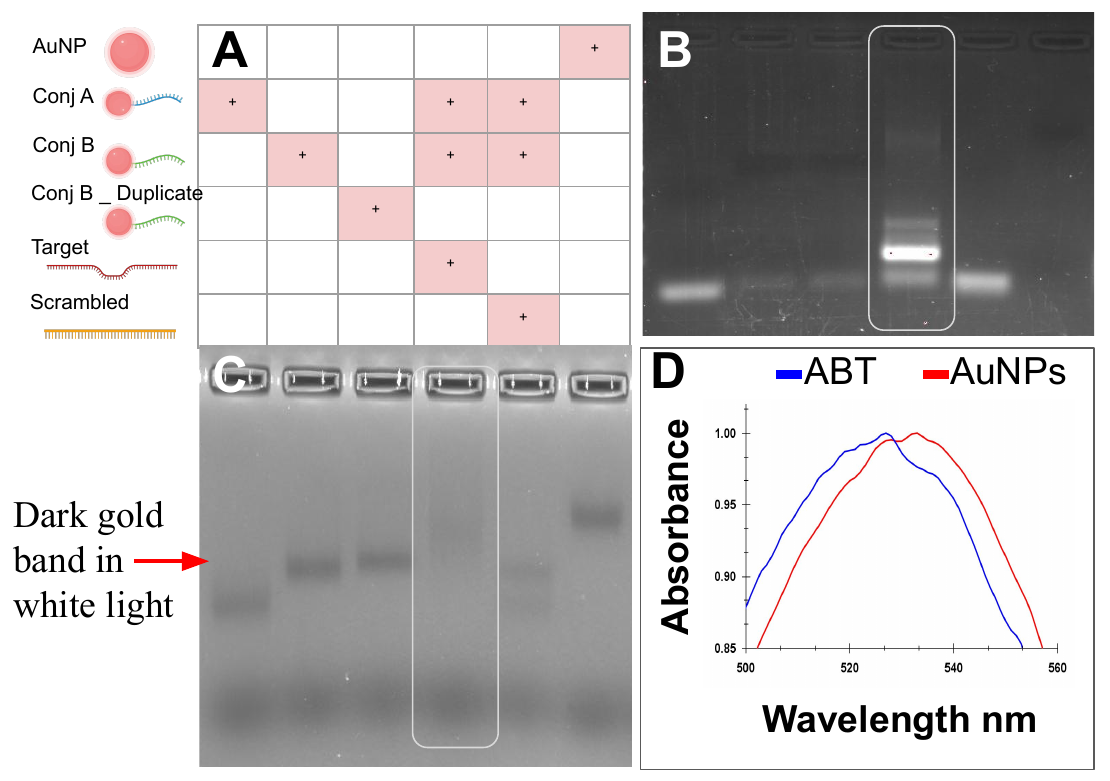}
\caption{\color{Gray} \textbf{Gel electrophoresis and nanospectrophotometry of gold nanoparticle conjugates mixed with target and scrambled DNA.} \textbf(A) Reaction mixtures loaded into the agarose gel for (B) ethidium bromide fluorescence and (C) epi white illumination images of a single gel. Dark bands in the epi white image represent gold nanoparticles, which co-register with a smeared fluorescence signal in the lane highlighted by the white rectangle, containing conjugates A, B, and the target. (D) Surface plasmon resonance peaks of the reaction mixture from conjugates A,B, and T  (blue line) versus bare gold nanoparticles (red line) collected by nanospectrophotometry.}

\label{fig3} 
\end{figure}

Presence of target DNA but not scrambled DNA shifted the electrophoretic mobility and surface plasmon resonance of gold nanoparticles-oligomer conjugates (Figure \ref{fig3}). Specific nanoparticle-oligomer reactants alone and mixed with target or scrambled DNA were loaded into agarose gel lanes, along with bare AuNPs as a control Figure \ref{fig3}A). Fluorescent bands of ethidium bromide staining indicate presence of double stranded (hybridized) DNA, strongest in a smeared band with the lowest electrophoretic mobility in the nanoparticle conjugated A-B-T lane Figure\ref{fig3}B). Moreover, epi white illumination revealed a smeared dark band representing light absorption by gold nanoparticles that co-registered with ethidium bromide signal, but no gold nanoparticle presence in the strongest fluorescent band (Figure\ref{fig3}C). The conjugated A-B-S mixture, in contrast, produced a double band of gold nanoparticle absorption with the same mobility as conjugate A and conjugate B alone, and with no co-registered fluorescence signal. The bare AuNPs with no DNA present had the least mobility. A ~6 nm blue shift in surface plasmon resonance was observed in the conjugated A-B-T mixture versus bare AuNP (peaks at 527 nm vs. 533 nm, (Figure \ref{fig3}D).

The addition of 2x and 3x stoichiometric ratios (6 $\mu M$ and 9 $\mu M$) of the target to oligomers A and B produced similar co-migrated smeared bands of gold absorption and ethidium fluorescence with proportionately slightly reduced mobility vs. the reaction with the target at 3 $\mu M$, 1x stoichiometric ratio (Figure \ref{figS2}A-C). Additional targets at 2x and 3x ratios had similar SPR as the 1x reaction mixture, with only ~1 nm red shift vs. 1x (peaks at 528 nm vs. 527 nm), still blue shifted from bare AuNPs with a peak at ~533 nm (Figure \ref{figS2}D).

\subsection*{4. Darkfield videomicroscopy}
\begin{figure}[H] 
\includegraphics[width=\textwidth]{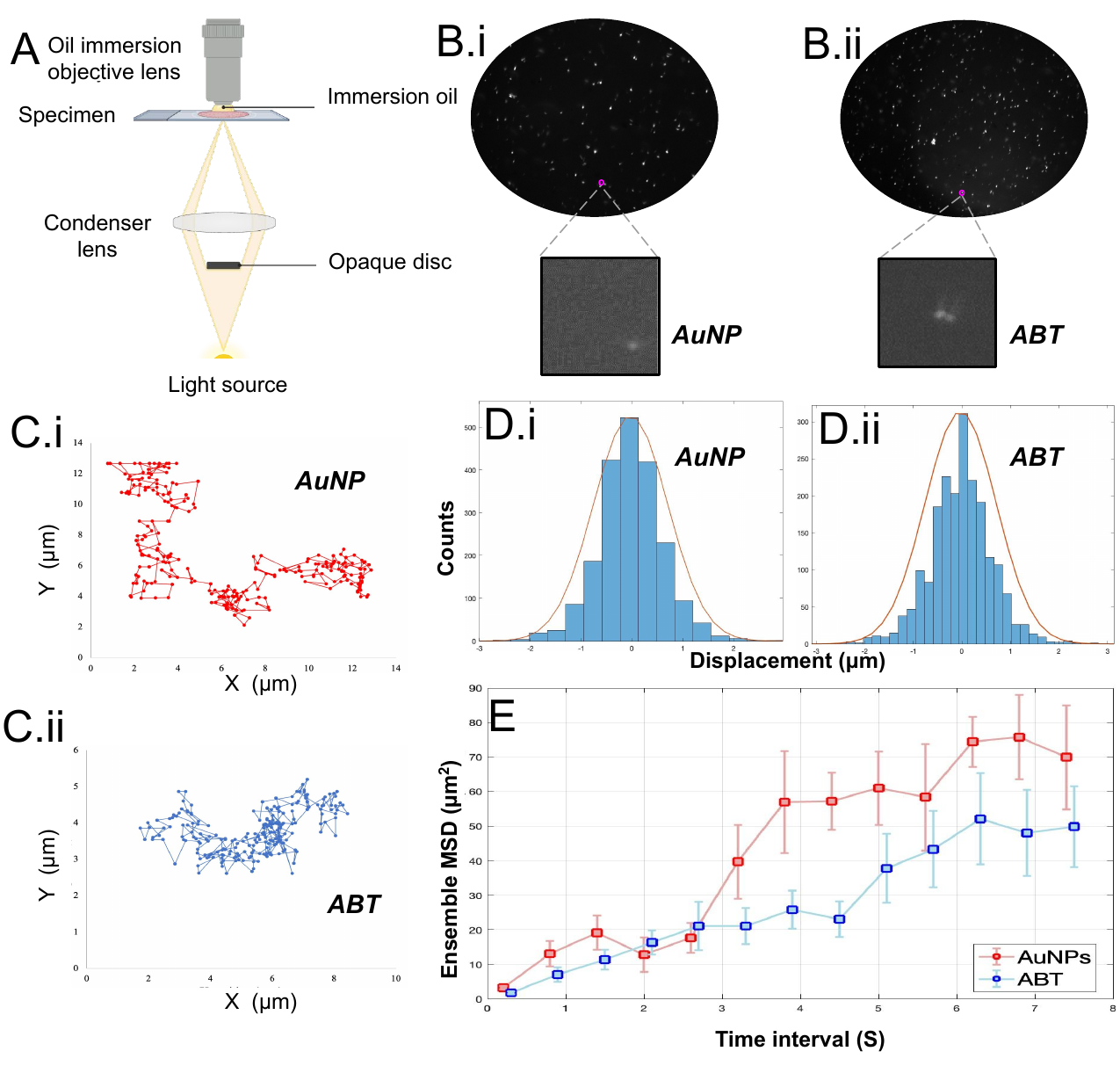}

\caption{\color{Gray} \textbf{Darkfield microscopy detection and nanoparticle tracking measurements of single gold nanoparticles and aggregates of gold nanoparticle conjugates with target DNA.} \textbf(A) Darkfield microscopy setup. (B) Raw darkfield micrographs of nanoparticle mixtures, (i) bare AuNPs and (ii) ABT, with representative nanoparticles for tracking, are depicted in the square zoom boxes. (C) Nanoparticle tracks over multiple darkfield frames for (i) bare AuNPs and (ii) ABT. Two methods of diffusion coefficient estimation from nanoparticle tracks were performed using (D) displacement histograms for (i) bare AuNPs and (ii) ABT and (E) mean squared displacement versus the time step for individual nanoparticle tracks from the bare AuNP mix (red) and the ABT mix (blue). Scale bars are indicated.}

\label{fig4} 
\end{figure}

Many diffusing scattering objects were detected in imaging chambers using darkfield microscopy (Figure \ref{fig4}A), both with bare AuNPs added (Figure \ref{fig4}B.i) and conjugates A and B with target (Figure \ref{fig4}B.ii). Nanoparticle tracking produced contiguous displacement tracks over 9 seconds and 300 frames for some of the objects in AuNP (Figure\ref{fig4}C.i) and A-B-T (Figure \ref{fig4}C.ii) mixtures. Gaussian fits to lateral displacement histograms were centered around 0 (Q1$\pm$R1, Q2$\pm$R2, and Q3$\pm$R3 mean$\pm$standard deviation), for AuNPs (Figure \ref{fig4}D.i), A-B-T (Figure\ref{fig4}D.ii), and A-B-S, respectively. Linear best fits of ensemble MSD vs. time interval were excellent (R\textsuperscript{2}=0.96$\pm$0.05, n=69 tracked objects, mean$\pm$standard deviation) (Figure \ref{fig4}E).

\section*{5. Diffusion coefficient estimation}
\begin{figure}[H] 
\includegraphics[width=\textwidth]{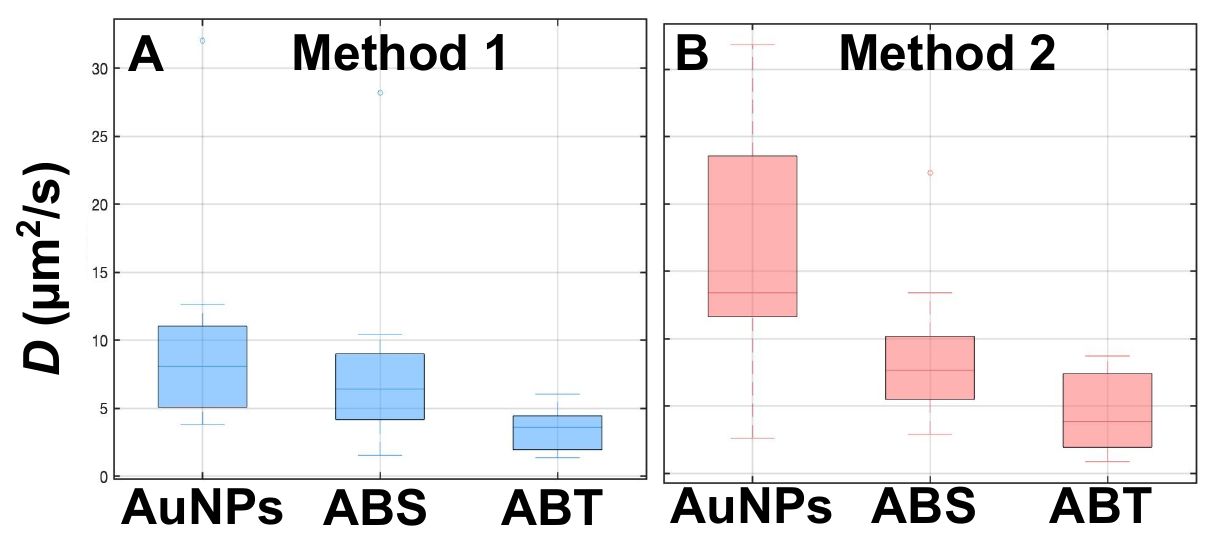}
\caption{\color{Gray} \textbf{Diffusion coefficient measurements from darkfield microscopy data.} \textbf Interquartile plots of estimated $D$ by (A) method 1 and (B) method 2 from particles tracked in bare AuNPs, ABS, and ABT mixtures.}
\label{fig5} 
\end{figure}

Diffusion coefficients estimated from dark-field microscopy were different for three nanoparticle mixtures (Figure\ref{fig5}). The mixture with AuNPs had the highest estimated diffusion coefficients by method 1 (Figure \ref{fig5}A), at 10.1$\,\pm\,$6.5 $\mu\text{m}^2/\text{s}$, followed by ABS at  7.3$\,\pm\,$5.4 $\mu\text{m}^2/\text{s}$ and ABT  3.6$\,\pm\,$1.5 $\mu\text{m}^2/\text{s}$. The type of mixture affected the estimated diffusion coefficient ($F=9.4$, $p<0.001$, 1-factor ANOVA). Specifically, A-B-T was different from AuNPs (Tukey test, $p<0.001$). Diffusion coefficients estimated by Method 2 showed a similar trend (Figure \ref{fig5}B). The mixture with AuNPs had the highest estimated diffusion coefficients at 36.5$\,\pm\,$38.0 $\mu\text{m}^2/\text{s}$, followed by ABS at 8.6$\,\pm\,$4.3 $\mu\text{m}^2/\text{s}$ and ABT at 4.8$\,\pm\,$3.2 $\mu\text{m}^2/\text{s}$. Mixture type significantly affected the estimated diffusion coefficient ($F=12.7$, $p<0.001$, 1-factor ANOVA). Pairwise comparisons showed that ABT and ABS were significantly different from AuNPs (Tukey test, $p<0.001$ for each).

\section*{6. Diffusion and Reaction Kinetics Simulations}
A stochastic simulation of spherical and cylindrical (aspect ratio 2:1) nanoparticle diffusion for a range of nanoparticle dimensions agreed well with theoretical predictions from Stokes-Einstein equations (Table 1). Cylindrical nanoparticles had 41±0.5 \% lower theoretical $D$ than spheres of equal radii, and 66±4.0 \% lower simulated $D$ than spheres of equal radii. The average error between $D$ calculated from Stokes-Einstein vs. measured from simulated data was 10.6±4.5 \%  for spheres and 40.5±6.5 \% for cylinders.

\begin{table}[H]
\centering
\resizebox{\textwidth}{!}{%
\begin{tabular}{|c|c|c|c|c|}
\hline
\textbf{Radius (nm)} & \textbf{\makecell{Stokes-Einstein,\\Sphere ($\mu$m$^2$/s)}} & \textbf{\makecell{Stokes-Einstein,\\Cylinder ($\mu$m$^2$/s)}} & \textbf{\makecell{Method 2,\\Sphere ($\mu$m$^2$/s)}} & \textbf{\makecell{Method 2,\\Cylinder ($\mu$m$^2$/s)}} \\ \hline
5 & 4.78 & 2.84 & 5.18 & 1.9 \\ \hline
10 & 2.4 & 1.42 & 2.55 & 0.8 \\ \hline
20 & 1.2 & 0.71 & 1.41 & 0.45 \\ \hline
50 & 0.48 & 0.28 & 0.42 & 0.17 \\ \hline
100 & 0.24 & 0.14 & 0.22 & 0.07 \\ \hline
\end{tabular}%
}
\caption{Comparison of diffusion coefficients estimated by different methods and for different nanoparticle shapes.}
\label{table:diffusion_coefficients}
\end{table}

\begin{figure}[H] 


\includegraphics[width=\textwidth]{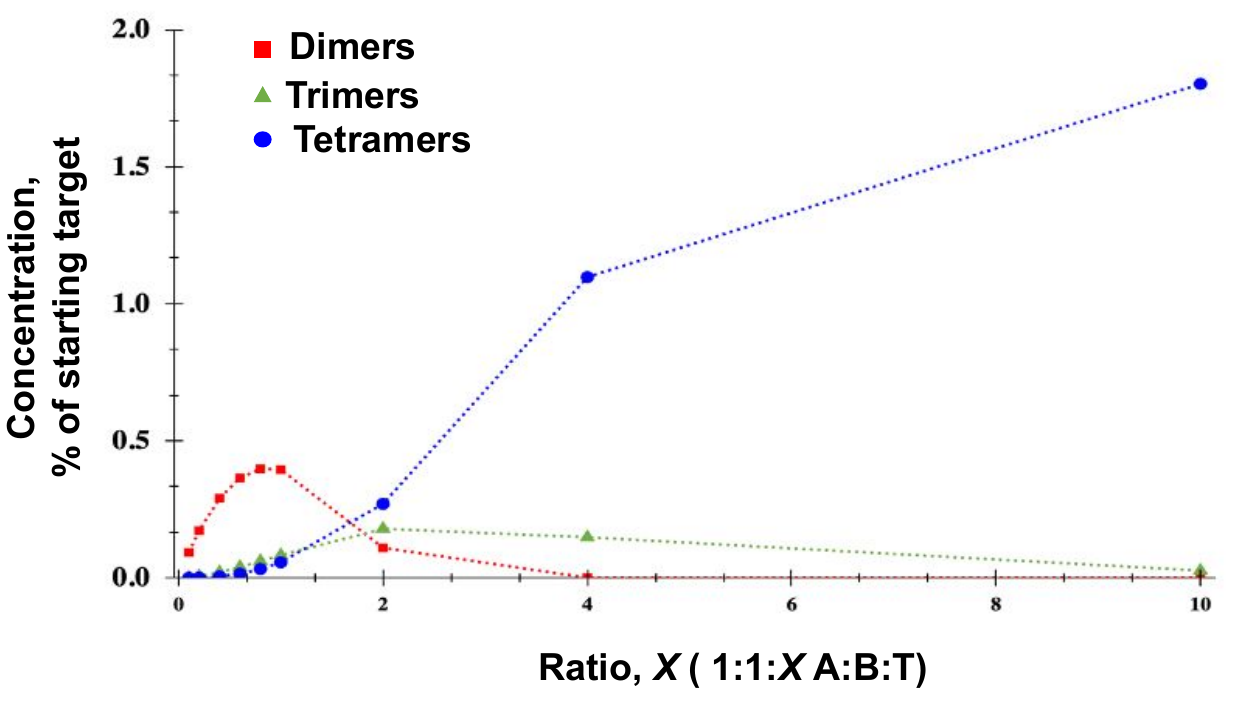}

\caption{\color{Gray} \textbf{Results of a kinetic nanoparticle-DNA hybridization reaction simulation.} \textbf The prevalence of dimer, trimer, and tetramer products, is expressed as a percent of starting target concentration (equals 100) versus the ratio $X$ of target to conjugates A and B.}
\label{fig6} 
\end{figure}

A stochastic reaction model utilizing first-order reaction kinetics found that the proportion of dimers, trimers, and tetramers in the reaction products depended on the initial stoichiometric ratio of target to oligomer-conjugated nanoparticles (Figure \ref{fig6}). At target: conjugate ratios $< 2$, dimers were predominantly formed. At a ratio of 2, dimers, trimers, and tetramers were all roughly equally present. At ratios of $<2$, tetramers were the predominant product formed. Higher-order aggregates were not modeled.

\section*{Discussion}
A nanoparticle aggregation assay was developed to detect the DNA sequence corresponding to a portion of the SARS-CoV-2 genome. Two oligomer-decorated nanoparticle populations allowing hybridization to target at the 5’ and 3’ ends were fabricated. While the hybridized trimer of 5’ oligomer, target, and 3’ oligomer formed with low free energy and no products from side reactions, the scrambled DNA sequence was able to mis-hybridize with only one of the oligomers. Nanoparticle-DNA aggregates were detected by gel electrophoresis and nanospectrophotometry in the presence of target but not scrambled sequence DNA. Nanoparticle tracking of these aggregates from darkfield videomicroscopy datasets determined that lower estimated diffusion coefficients occurred in the presence of the target than with the scrambled sequence or no added DNA. Further, the methods of diffusion coefficient estimation were reasonably accurate and able to distinguish between single nanoparticles, dimers, trimers, and tetramers, modeled as both spheres and as cylinders. Kinetic modeling suggested that tetramers predominate when the target is present at a more than 2:1 ratio with nanoparticle conjugates, and dimers predominate when the target is present at <1:1. This for low levels of target the assay would detect almost exclusively dimers, with more aggregates as target molecules become more abundant.
The effective development of a reliable method to sensitively detect DNA sequences without amplification from SARS-CoV-2, or any source of nucleic acids, depends on several crucial elements, namely either the ability to measure many targets binding events A) together in an ensemble, or else B) individually but separated in space and time. Detection methods that rely upon an ensemble of target binding molecules acting together include shifts in electrophoretic mobility of DNA and gold nanoparticles, with corresponding fluorescence and absorption signals that co-migrate, and a shift in surface plasmon resonance detected by nanospectrophotometry. Both methods are limited by the instrument capabilities: for electrophoresis, the pixel dimensions (resolution) and sensitivity of the camera in the gel imager, and for nanospectrophotometry, the sensitivity of the photodetector in terms of spectral linewidth and change in absorbance. In contrast, target detection by nanoparticle aggregation depending on nanoparticle tracking with darkfield videomicroscopy allows individual binding event detection and detects many particles in parallel spatially across the camera sensor, and serially over time.  While 10 nm diameter gold nanoparticles are close to the limit of sensitive detection by darkfield microscopy, aggregation ensures dimers and larger aggregates formed upon target binding should be both brighter and slower, and thus easier to detect. In that case, the sensitivity of detecting nanoparticle bound target DNA depends on the number of nano-sized objects able to be detected in the microscope field of view for the data acquisition period. With an automated microscope, roughly five 9-second videomicroscopy datasets could be collected per minute by shifting the field of view and automating acquisition. On the other hand, a lack of monodisperse nanoparticles in the starting mixture will tend to obscure nanoparticle aggregation due to target binding. In that case, statistical analyses of sampled populations of target/unknown versus control (no target) specimens is required to allow confident detection of a given target DNA sequence.
A key feature distinguishing the proposed DNA detection approach is the tracking of diffusing nanoparticles and nanoparticle aggregates. Previously, nano- and micro-spheres were used to bind DNA target sequences, and were detected by light microscopy above the diffraction limit, or else with transmission electron microscopy, for example in immunogold labeling. The advantages of detecting diffusing nanoparticles include that more nanoparticles can be detected per image frame and over different fields of view imaged serially, or else the same field of view but with the reaction mixture replaced by fluid flow. Thus in the future, nanoparticle tracking may be performed in a microfluidic device. A microfluidic platform would have the advantage of the ability to test a small specimen volume and potentially complete coverage of the microfluidic channel spanned by one field of view. A 50-micron tall channel, though, would only assay about 8 nL, so a 2 $\mu L$ specimen volume would have to be displaced and repeatedly measured 250 times to measure the entire specimen. Channel splitting and image multiplexing would reduce the time to acquire such a complete dataset.
In conclusion, the results described above demonstrate the feasibility of detecting specific DNA sequences by tracking diffusing nanoparticles and measuring their diffusion coefficients. With added features, darkfield microscopes can also sensitively detect shifts in the spectrum, brightness, and polarization of light scattered from nanoparticles. This suggests that in the future, multiple optical signals could be collected at the same time in a single dataset to refine the sensitivity and specificity of target detection. This combined with microfluidic platforms and automated videomicroscopy data collection may allow total analysis of a small reaction volume (2 $\mu L$ or so) in a reasonably short amount of time, and without signal amplification by PCR, enzymatic activity or by ensemble detection. These findings support further development of microscopy based biomolecular detection assays.

\section*{Supporting Information}
\section*{Design and Optimization of Primers}
The selection of target sequences was meticulously executed to ensure a precise and efficient amplification of the RNA-dependent RNA polymerase (RdRp) region of the SARS-CoV-2 genome (GenBank accession number MT072668.1). Using PrimerBlast (NCBI), we initially generated 10 primers from the RdRp region within ORF1ab of the SARS-CoV-2 genome, each spanning 30 nucleotides. Oligo A (forward) corresponded to gene position 9375–9404, while Oligo B (Reverse of Reverse Complement) spanned gene position 9447–9418.
The primer design adhered to stringent criteria that were aimed at optimizing functionality and PCR efficiency. The primers were designed with a GC content of 40–60\%, primer ends containing at least two guanine or cytosine nucleotides, a primer length of at least 18 base pairs and no complementary sequences between the two primers. Additionally, the melting temperatures (Tm) of the forward and reverse primers were maintained between 55 and 65°C  \cite{thornton2015}. Additionally, complementary sequences between forward and reverse primers were meticulously avoided to prevent primer-dimer formation and nonspecific amplification. The reverse complement sequences of the initial primers were computed, as these sequences would hybridize with cDNA reverse transcribed from the original viral RNA \cite{sen2017}. 
\section*{Target Selection and Scrambled DNA Preparation}
In order to identify the sequences accurately we specifically selected them to ensure amplification of acid. We considered factors such, as the uniqueness of the sequence the size of the amplicon and the accessibility of the target area when choosing the targets, for detecting the RdRp segment of SARS-CoV-2 \cite{yu2020}. Target DNA sequences were matched with bead capture probe combinations. Underwent a process known as hybridization. To prevent stress 13 nucleotides was strategically inserted to disrupt the bonding, between the two oligonucleotides and the target.
Preparation of Scrambled DNA Controls: Scrambled DNA controls were meticulously prepared to serve as negative controls in PCR assays. These controls were designed to mimic target sequences' length and GC content while containing randomized nucleotide sequences. By incorporating scrambled DNA controls, the specificity and reliability of PCR assays were validated, thereby enhancing the accuracy of molecular analyses. 

\begin{figure}[H] 
    \centering 
    \includegraphics[width=\textwidth]{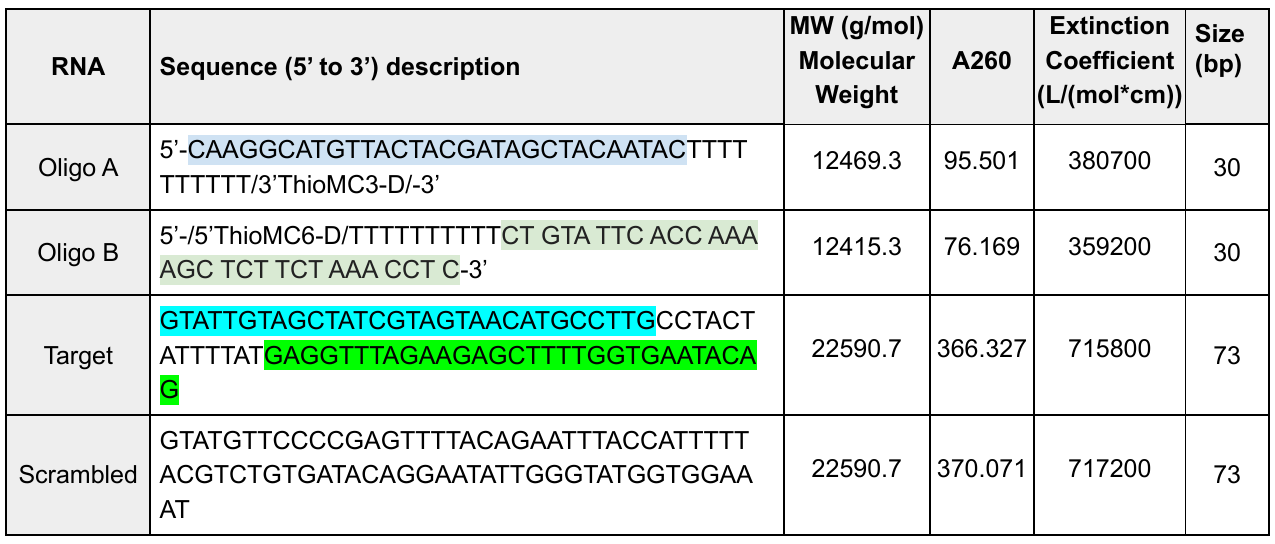} 
    \caption{\textbf{Table S1} Table of oligo, target, and scrambled sequences with their corresponding molecular weights, extinction coefficients, and sizes.}
    \label{TableS1} 
\end{figure}

\subsection*{Figure S1}
\begin{figure}[H] 

\renewcommand{\thefigure}{S1}

\includegraphics[width=\textwidth]{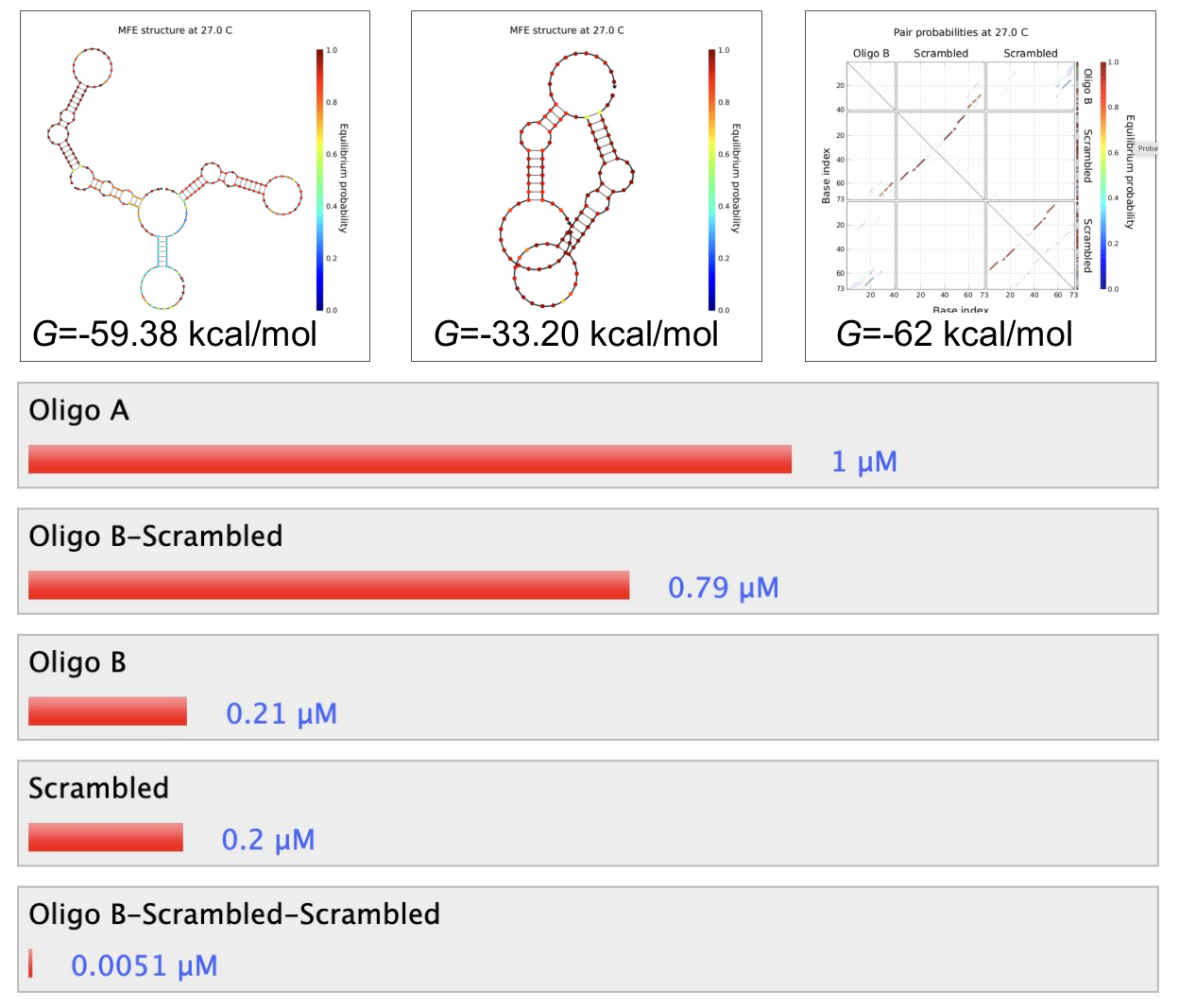}

\caption{{\bf Nupack simulation of hybridization rate and structure of Oligo A, Oligo B with scrambled.} The simulations yielded two distinct secondary structures with Oligo B, with free energy values of -59.38 kcal/mol and -33.20 kcal/mol, respectively. However, the second structure of -33 kcal/mol is noteworthy in that the structure was deemed unrealistic and difficult to form under physiological conditions. The simulations suggest that there is no feasible hybridization potential between Oligo A and the scrambled sequence. These findings underscore the specificity of the hybridization process and highlight the selective binding nature of oligonucleotide sequences.}
\label{figS1} 
\end{figure}

\subsection*{Figure S2}
\begin{figure}[H] 

\renewcommand{\thefigure}{S2}

\includegraphics[width=\textwidth]{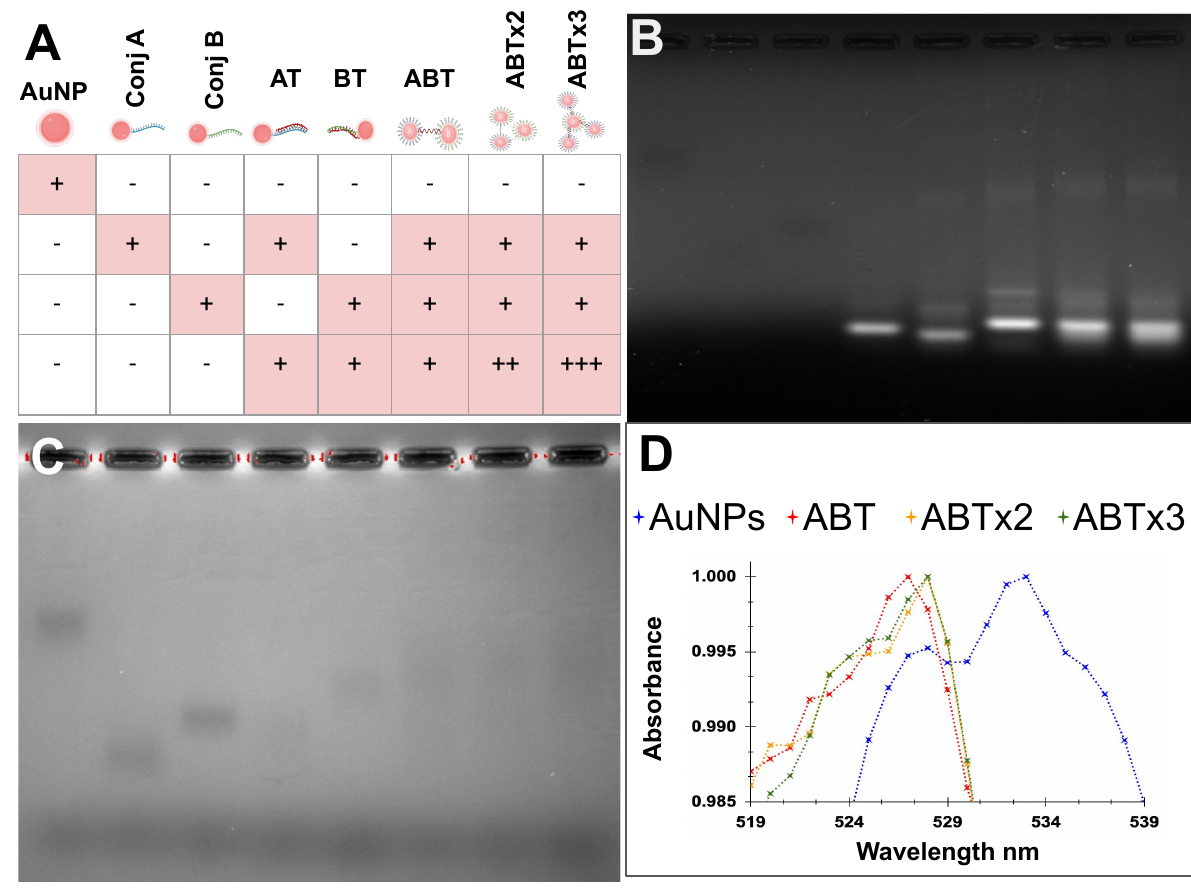}

\caption{{\bf Schematic Representation of Target-Induced Aggregation of Conjugate A and Conjugate B.} . (B) The schematic representation shows the ethidium bromide gel image displaying dimers, trimers, and tetramers formed with increasing target concentrations. (C) Gel images depict the conjugates A and B and their respective aggregates formed upon hybridization with increasing target concentrations. (D) UV-vis spectroscopy data demonstrates the shifting of the peak at different concentrations of the target to Conjugate A and Conjugate B, indicating the aggregation process.}
\label{figS2} 
\end{figure}

\subsection*{Figure S3}
\begin{figure}[H] 
\renewcommand{\thefigure}{S3}
\centering
\begin{minipage}{0.5\textwidth} 
    \centering
    \includegraphics[width=\textwidth]{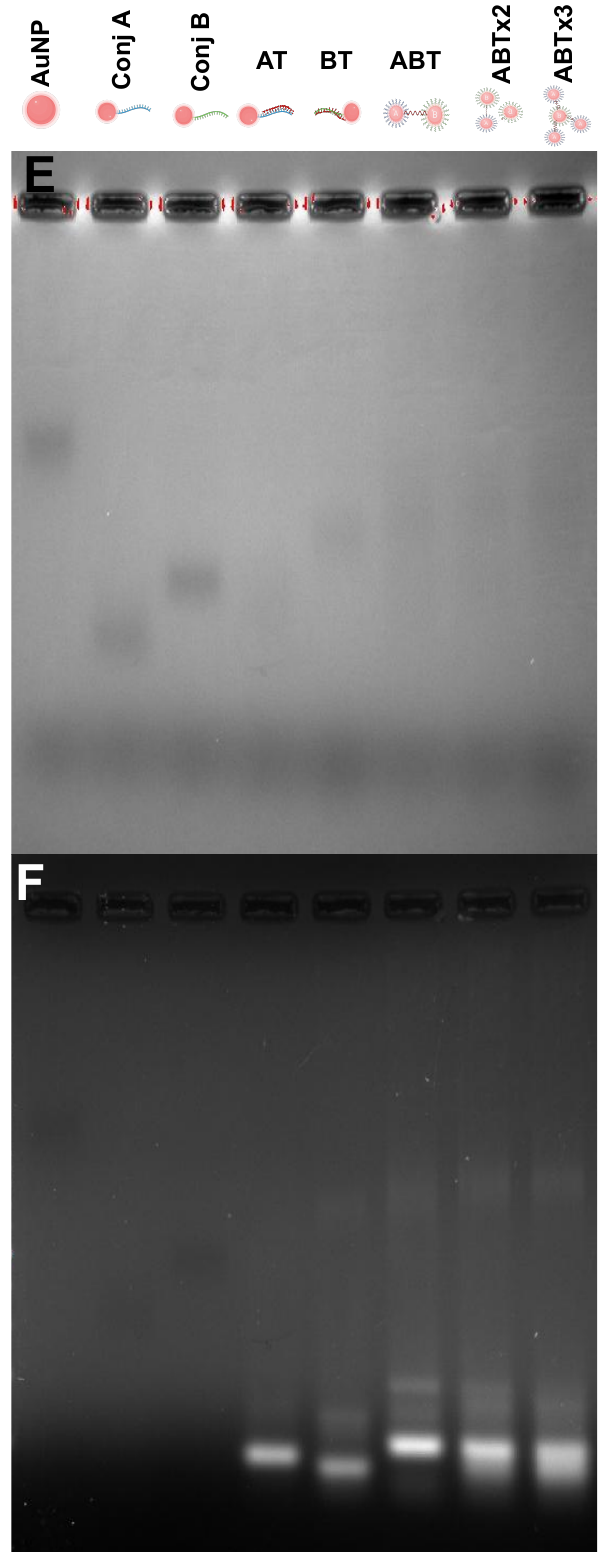} 
\end{minipage}%
\begin{minipage}{0.45\textwidth} 
    \caption{{\bf Process of target-induced aggregation of Conjugate A and Conjugate B, represented in monovalent, divalent, and trivalent forms, corresponding to one, two, and three times the target concentration.} (A) Gel images depict the conjugates A and B and their respective aggregates formed upon hybridization with increasing target concentrations. (B) The schematic representation shows the ethidium bromide gel image displaying dimers, trimers, and tetramers formed with increasing target concentrations.}
    \label{figS3} 
    \section*{Diffusion Motion Collection}
    Using a Nikon ECLIPSE TE300 microscope to image gold nanoparticles and DNA-AuNPs samples were diluted in an aqueous solution and captured over time. By placing a piece of double-sided imaging spacer (0.15 mm) between two coverslips (0.13-0.17 mm), the sample thickness of 0.5 mm was attained. Sealed imaging spacers were used to prevent evaporation and fluid flow during imaging. Tracking Fiji image processing software was used to extract particle positions (Fig. 4C.i-ii) \cite{schindelin2012}.
    \section*{Fiji image processing software}
    The examination of diffusion states was performed utilizing the TrackMate plugin, a component of the Fiji image processing software that facilitates the tracking of individual particles. With the TrackMate program, spots or roughly spherical objects can be partially automatically separated from both 2D images. This lets you track them over a certain amount of time. For the detection methods, the LoG (Laplacian of Gaussian) segmentation method is used. The mobile diffusions of the samples were classified by utilizing a diffusion coefficient threshold, which was determined based on the dynamic localization computed for each individual cell.
\end{minipage}
\end{figure}

\subsection*{Video S1}
Movie of AuNPs diffusion and darkfield video. You can view the video online at \url{https://github.com/Aldakheelarwa88/AuNPs_Diffusion_Videos.git}.

\subsection*{Video S2}
Movie of ABS diffusion and darkfield video. You can view the video online at \url{https://github.com/Aldakheelarwa88/ABS_Diffusion_Videos.git}.

\subsection*{Video S3}
Movie of ABT diffusion and darkfield video. You can view the video online at \url{https://github.com/Aldakheelarwa88/ABT_Diffusion_Videos.git}.

\section*{Acknowledgments}
We would like to express our sincere gratitude to all those who contributed to the success of this work. Special thanks to our colleagues for their invaluable insights and discussions, and to our advisors for their continued guidance and support. We extend our appreciation to Dr. Xiaolong Luo, Juewen Liu, and Georges Nehmetallah for their support and valuable contributions. Our heartfelt thanks go to our colleagues for their invaluable insights and discussions, and to the laboratory staff for their technical assistance. We are equally grateful to our collaborators for their contributions to this research. Lastly, we are thankful to our funding agencies SACM for their financial support, which made this project possible.

\newpage

\bibliography{library} 

\bibliographystyle{unsrt} 

\newpage


\end{document}